# Nonlinear Dynamics of DNA Chain

Deoxyribonucleic acid (DNA) is certainly one of the most interesting molecules. Interest in its structure and dynamics is primarily due to the important role that this molecule plays in life processes. The molecule was first identified in the 1860s by a Swiss chemist Johann Friedrich Miescher. He discovered a substance that had unexpected properties, different from those of the other proteins he had been familiar with. He did not know he had discovered the molecular basis of life called by him nuclein.

Great progress was made by a German biochemist Albrecht Kossel who identified the nuclein as a nucleic acid in 1881. He isolated the four nucleotide bases that are the building blocks of DNA, introduced their present names and obtained the Nobel Prize in 1910.

A revolution has been related to a famous Watson-Crick model, published in one-page paper [1]. The paper was published in 1953 and the authors were awarded the Nobel Prize in Medicine in 1962. According to the model, DNA is a double helix, formed from two mutually complementary strands, as shown in Fig. 1. We assume that the readers have basic knowledge about its structure [2-7]. It suffices now to point out that each strand represents a series of nucleotides, whose constituent parts are sugar, phosphate, and base. The nucleotides are always linked together by strong covalent bonds, while different strands interact through basis by weak hydrogen bonds. Adenine (A) is always attached to thymine (T) by two hydrogen bonds, whereas guanine (G) and cytosine (C) are attached by three bonds. This is shown in Fig. 1, where we recognize two strands representing sugar-phosphate backbones and four kinds of basis. All sugars and phosphates are equal, which means that genetic information depends on the basis only.

In addition to H-bonds, the stability of the chain is supported by stacking interactions [5,8]. These weak forces, interacting between neighboring bases of the same chain, are of crucial importance for DNA twisting determination.

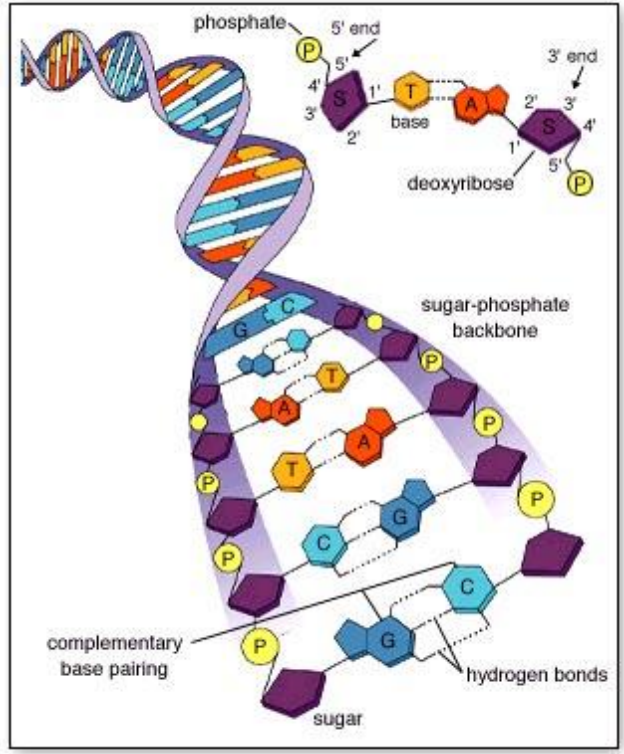

Figure 1: The structure of DNA molecule

## 1. DNA Dynamics

To date, dozens of different mechanical models and their versions have been developed to describe DNA dynamics. For the simplest structural model DNA chain is an elastic rod [9]. More advanced models are helical double rod-like models [9]. In both cases, the rods can be either uniform or discrete. According to these simple models, plain waves propagate along the chain.

It was in 1980 when Englander et al. suggested that nonlinear effects might play an important role in the DNA dynamics [10]. Instead of the plain waves, the nonlinear effects may focus the vibration energy of DNA into localized soliton-like excitations. Therefore, we can talk of linear and nonlinear models. This is very important to be understood and requires some explanations.

Suppose that there is strong interaction between two neighboring nucleotides. An example can be the covalent bond between the nucleotides belonging to the same strand, as explained above. The existence of the strong force means that displacements along the direction of this force are very small. In other words, the oscillations in this direction have small amplitudes. This means that we can assume that attractive and repulsive forces are almost equal and the corresponding potential energy, or potential for short, should be modeled by a symmetric function. A typical example is the well-known function $f(x) = kx^2/2$. Such potential is called harmonic and its usage in science has been called harmonic approximation. Its first derivative represents a force, which is obviously a linear function. On the other hand, if these forces are weak, the corresponding displacements are big and the repulsive and attractive forces are not equal any more. An example can be the hydrogen bond between the nucleotides belonging to different strands, which was mentioned above. Therefore, to model such potentials we need non-symmetric functions. A common example is the function $F(x) = D[e^{-ax} - 1]^2$, shown in Fig. 2. This potential energy is called Morse potential. The parameters $D$ and $a$ are the depth and the inverse width of the Morse potential well, respectively. The first derivatives for negative and positive $x$ represent the repulsive and attractive forces, respectively, and it is obvious that the latter one is smaller. For very big distance between the interacting particles the first derivative is zero, which means that the particles do not interact any more. This potential is not harmonic and its first derivative, i.e., the force, is not the linear function. The models that include at least one anharmonic interaction are called nonlinear. Therefore, the weak interactions are the sources for the nonlinear terms and, consequently, such systems are nonlinear. As these weak forces are common for biological systems, we concentrate on the nonlinear models only.

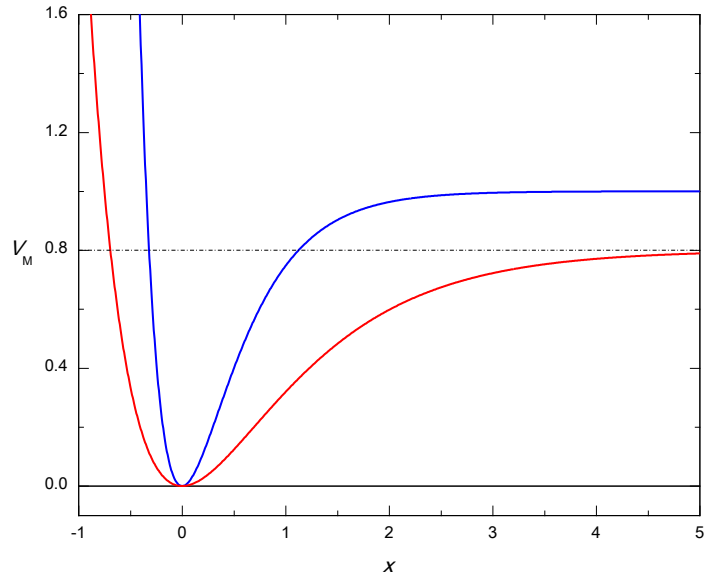

Figure 2: Morse potential energy $F(x) = D[e^{-ax} - 1]^2$ for $D = 1$, $a = 2$ (blue) and $D = 0.8$, $a = 1$ (red)

Let us explain the first nonlinear model. According to the model, DNA represents two linear chains of pendulums (the bases) connected to the sugar-phosphate backbones [10]. If $\theta_n$ is the angle between the pendulum and direction around which it oscillates then the total energy is given by the following Hamiltonian

$$H = \sum \left[ \frac{mh^2}{2} \left( \frac{d\theta_n}{dt} \right)^2 + \frac{S}{2}(\theta_n - \theta_{n-1})^2 + mgh(1 - \cos\theta_n) \right], \qquad (1)$$

where $m$ and $h$ are mass and length of the pendulum, respectively [10]. It is clear that nonlinearity is coming from the cosine function.

Using Hamilton's equations and appropriate generalized coordinates, i.e., $\dot{q}_n = \partial H/\partial p_n$, $\dot{p}_n = -\partial H/\partial q_n$, $q_n = \theta_n$ and $p_n = mh^2\dot{\theta}_n$, where the dot means the first derivative with respect to time, the Hamiltonian (1) brings about an appropriate equation of motion, that is

$$mh^2\left(\frac{d^2\theta_n}{dt^2}\right) = S(\theta_{n+1} + \theta_{n-1} - 2\theta_n) - mgh\sin\theta_n, \quad (2)$$

where $S$ is a harmonic constant, while the number $n$ determines the position of the pendulum. In the static limit, when the term including the second derivative is neglected [10], we obtain the equation

$$S(\theta_{n+1} + \theta_{n-1} - 2\theta_n) - mgh\sin\theta_n = 0 \quad (3)$$

whose solution is

$$\theta_n = 4\arctan[\exp(2nl/L)], \quad L = 2l\sqrt{S/mgh}, \quad (4)$$

where $l$ is the distance between two neighbouring pendulums [10]. The function $\theta_n \equiv \theta(n)$ is a kink soliton.

It is convenient, very often, to begin with the discrete case and then pass to the continuum limit, which will be followed here. This means that, when $\theta_n$ does not vary too rapidly with $n$, the following series expansion can be performed

$$\theta_{n\pm1}(t) \approx \theta(z,t) \pm \theta_z(z,t)l + \frac{1}{2}\theta_{zz}(z,t)l^2 \quad (5)$$

and Eq. (2) becomes

$$mh^2\left(\frac{d^2\theta_n}{dt^2}\right) - S\left(\frac{d^2\theta_n}{dx^2}\right) + mgh\sin\theta_n = 0. \quad (6)$$

This is well-known solvable sine-Gordon equation [11,12]. Its solutions are the kink and antikink solitons.

The procedure explained above can be extended and the DNA molecule can be seen as a series of coupled double pendulums [13,14]. The first pendulum models the oscillation of the phosphate-sugar part of nucleotide, while the remaining one describes the oscillation of the base. This approach can be further extended in order to describe inhomogeneous DNA chains [15].

Now, we explain Y-model, introduced by Yakushevich in 1989 [16]. According to the model, DNA consists of two parallel chains of discs. The chains are straight, which means that the helicoidal structure is not taken into consideration. The discs are connected with each other with longitudinal and transverse springs. The rigidity of the longitudinal springs is higher than that of the transverse ones as they represent the covalent and hydrogen bonds, respectively. Let us suppose that the chains are in z-direction, while the surfaces of the discs are in xy-plain. The model assumes angular oscillations of the discs in the xy-plain, only. This means that the Hamiltonian of DNA is [16]

$$H = \sum_{i,n}\left[\frac{I_i}{2}\dot\varphi_{i,n}^{\,2} + \frac{K_i}{2}(\varphi_{i,n}-\varphi_{i,n-1})^2 + \frac{k}{2}(\Delta l_n)^2\right], \qquad (7)$$

where $i=1,2$ and $n=1,2,...$ denote the chains and discs, respectively, $K_i$ and $k$ are the rigidities and $\Delta l_n$ is the stretching of the nth transverse spring due to rotations of the discs. Let us determine $\Delta l_n$ first. Imagine two discs of the radius $R$ in the same plain at the position $n$. Then, the distance between their centers is $2R+l_0$, where $l_0$ is nothing but the length of the unstretched spring. Let us imagine that these two discs perform the angular displacements $\varphi_{1,n}$ and $\varphi_{2,n}$. The new length becomes $l_n$, where

$$l_n^{\,2} = (l_0 + 2R - R\cos\varphi_{1,n} - R\cos\varphi_{2,n})^2 + (R\sin\varphi_{1,n} - R\sin\varphi_{2,n})^2, \qquad (8)$$

which brings about

$$\Delta l_n = l_n - l_0. \qquad (9)$$

Using the Hamilton's equations and the generalized coordinates $q_i = \varphi_i$ and $p_i = I_i\dot\varphi_i$, we easily obtain the dynamical equations of motion according to Eqs. (7)-(9). The one for $\varphi_{1,n}$ is

$$I_1\ddot\varphi_{1,n} = K_1(\varphi_{1,n+1}+\varphi_{1,n-1}-2\varphi_{1,n})^2 - k\frac{\Delta l_n}{l_n}\left[(2R^2+Rl_0)\sin\varphi_{1,n} - R^2\sin(\varphi_{1,n}+\varphi_{2,n})\right], \qquad (10)$$

while the remaining equation can be obtained from Eq. (10) by replacing the index 1 by 2 and vice versa.

The next step is the continuum limit. This means that we replace $\varphi_{i,n}(t)$ with $\varphi_i(z,t)$, where the coordinate $z$ is in the direction of the chains, as explained above. This simplifies Eq. (10) and we straightforwardly obtain

$$I_1\ddot\varphi_1 = K_1 a^2 \varphi_{1zz} - k\frac{\Delta l}{l}\left[(2R^2+Rl_0)\sin\varphi_1 - R^2\sin(\varphi_1+\varphi_2)\right], \qquad (11)$$

as well as the corresponding one for $\varphi_1 \leftrightarrow \varphi_2$. Here, the index $z$ represents the derivative with respect to the coordinate $z$. Notice that $\Delta l/l$ depends on the functions $\varphi_1$ and $\varphi_2$, which means that Eq. (11) is very far from being solvable. Fortunately, we can assume $l_0 << R$, which suggests a new approximation $l_0 \approx 0$ [16], yielding to $\Delta l/l = 1$ in Eq. (11). All this brings about the following system of equations

$$I_1\ddot{\varphi}_1 = K_1 a^2 \varphi_{1zz} - kR^2[2\sin\varphi_1 - \sin(\varphi_1 + \varphi_2)],$$
$$I_2\ddot{\varphi}_2 = K_2 a^2 \varphi_{2zz} - kR^2[2\sin\varphi_2 - \sin(\varphi_1 + \varphi_2)].$$
(12)

This system is nonintegrable in general. In the case of linear approximation, i.e., for very small oscillations, the solutions for $\varphi_1$ and $\varphi_2$ are plain waves, as expected. For $\varphi_1 = -\varphi_2$ and $\varphi_1 = \varphi_2$, Eqs. (12) reduce to the sine-Gordon and double sine-Gordon equations, respectively. These particular equations have one feature in common: they have the soliton-like solutions named kinks and antikinks. So, we can expect that the system (12) might also have the soliton-like solutions of kink and anti-kink type [16].

Y-model has been subjected to variety of improvements. For example, helicoidal structure of DNA was taken into consideration in [17], while in [18] it was not assumed that all bases are equal and, consequently, more realistic model was explained. In [19] the author did not assume the approximation $l_0 \approx 0$, while Morse potential was introduced in [20] to model the weak hydrogen bonds. Of special importance is the composite Y-model [21]. A key point is that the sugar-phosphate group and base are described by separate degrees of freedom. The composite Y-model contains the Y-model as a particular case. It represents an improvement providing a more realistic description of DNA. We should point out that the existence of solitons is a generic feature of the underlying nonlinear dynamics and is to a large extent independent of the detailed modeling of DNA [21]. Finally, let us point out that the Y-model allows us to study DNA dynamics under external influences [20,22,23].

A key problem in each model is a choice of degrees of freedom. Namely, DNA dynamics can be connected with either angular or radial displacements of the bases from their equilibrium positions. This means that, if we assume only one degree of freedom per nucleotide, we can choose either angular or radial variable as coordinate and the appropriate models can be called angular (torsional) and radial models, respectively. Of course, some extensions, i.e., models combining both kinds of coordinates, are possible [24-29]. The models mentioned above [10,16,21] are obviously the angular ones. In what follows, we describe a couple of radial models. We start with the Peyrard-Bishop (PB) model and further describe its two improvements, which we call the helicoidal Peyrard-Bishop (HPB) and Peyrard-Bishop-Dauxois (PBD) models.

To understand the PB model, we should remind ourselves of the chemical bonds existing between the nucleotides within DNA. In Fig. 3, we recognize the two strands. The interactions between the nucleotides belonging to the same strand are very strong and the corresponding oscillations are negligible. On the other hand, the bases of the nucleotides belonging to the different strands, interact through the weak hydrogen bonds, modeled by the Morse potential, as explained above. Hence, these oscillations are not negligible and it certainly makes sense to choose the radial displacements $u_n$ and $v_n$ as the crucial degrees of freedom. Notice that the strands are the linear systems, while DNA, due to the weak interactions, is not.

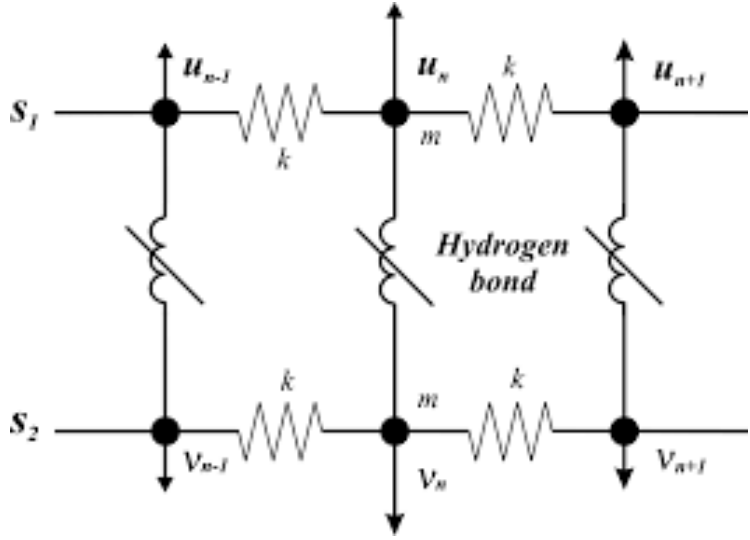

Figure 3: A simplified structure of DNA molecule

Like above, we start with the discrete Hamiltonian and pass to the continuum limit. According to the PB model, the Hamiltonian for DNA, in the nearest neighbor approximation [30,31], is

$$H = \sum \left\{ \frac{m}{2}(\dot{u}_n^2 + \dot{v}_n^2) + \frac{k}{2}[(u_n - u_{n-1})^2 + (v_n - v_{n-1})^2] + D[e^{-a(u_n - v_n)} - 1]^2 \right\}, \qquad (13)$$

where $m = 300\,\text{amu} = 5.1 \cdot 10^{-25}\,\text{kg}$ is the nucleotide mass, $k$ is a constant of the harmonic interaction and $\dot{u}_n$ and $\dot{v}_n$ represent the appropriate velocities. Obviously, the coordinates $u_n$ and $v_n$ are longitudinal (radial) displacements of the nucleotides at the position $n$ from their equilibrium positions along the direction of the hydrogen bond. One can recognize the kinetic energy term, potential energy describing the covalent interaction and Morse potential.

It is convenient to introduce new coordinates representing the in-phase and the out-of-phase transversal motions as

$$x_n = (u_n + v_n)/\sqrt{2}, \qquad y_n = (u_n - v_n)/\sqrt{2}, \qquad (14)$$

which transforms Hamiltonian (13) into

$$H = \sum \left\{ \frac{m}{2}(\dot{x}_n^2 + \dot{y}_n^2) + \frac{k}{2}[(x_n - x_{n-1})^2 + (y_n - y_{n-1})^2] + D[e^{-a\sqrt{2}y_n} - 1]^2 \right\}. \qquad (15)$$

Therefore, the coordinate $x_n$ describes the oscillation of the center of mass, while $y_n$ is proportional to stretching of the nucleotide pair at the position n. We are going to see that the function $x_n(t)$ represents a linear wave, while $y_n(t)$ is a nonlinear one and, consequently, more interesting for us. It is important to know that $y_n(t)$ is not temperature dependent but its mean value $\langle y \rangle$ is [30-32]. We do not explain the model in details as this will be done through one of its improved versions. In fact, the PB model is a special case of the HPB model.

The PB model does not take helicoidal DNA structure into consideration, while its improved version, the HPB model, does. This has been achieved introducing an additional term, describing helicoidal interactions, into Eq. (3.15) and the Hamiltonian becomes [33]

$$H = \sum \left\{ \frac{m}{2}(\dot{u}_n^2 + \dot{v}_n^2) + \frac{k}{2}[(u_n - u_{n-1})^2 + (v_n - v_{n-1})^2] \right. $$
$$\left. + \frac{K}{2}[(u_n - v_{n+h})^2 + (u_n - v_{n-h})^2] + D[e^{-a(u_n - v_n)} - 1]^2 \right\}, \quad (16)$$

where $K$ is the harmonic constant of the helicoidal spring. To understand the new helicoidal term we should imagine DNA in Fig. 3 being twisted. This means that after a turn of $\pi$, the nucleotide belonging to one strand at the position $n$ will be close to both ($n+h$)-th and ($n-h$)-th nucleotides of the other strand [33]. As the helix has a helical pitch of about 10 base pairs per turn [34] one can assume $h = 5$. Of course, we use Eq. (14) again, as well as the generalised coordinates $q_{nx} = x_n$, $q_{ny} = y_n$, $p_{nx} = m\dot{x}_n$, $p_{ny} = m\dot{y}_n$ and obtain the following completely decoupled dynamical equations of motion

$$m\ddot{x}_n = k(x_{n+1} + x_{n-1} - 2x_n) + K(x_{n+h} + x_{n-h} - 2x_n) \quad (17)$$
$$m\ddot{y}_n = k(y_{n+1} + y_{n-1} - 2y_n) - K(y_{n+h} + y_{n-h} + 2y_n) + 2\sqrt{2}aD(e^{-a\sqrt{2}y_n} - 1)e^{-a\sqrt{2}y_n}. \quad (18)$$

In the continuum limit, i.e., if we apply Eq. (5), both terms in the brackets in Eq. (17) will transform into the second derivatives with respect to the spatial coordinate. This means that we obtain an ordinary wave equation whose solution is the usual linear wave (phonon). However, Eq. (18) describes nonlinear waves. We restrict our attention on it and assume that the oscillations of nucleotides are large enough to be anharmonic but still small enough so that the nucleotides oscillate around the bottom of the Morse potential well. This suggests the transformation

$$y_n = \varepsilon \Phi_n; \qquad (\varepsilon \ll 1) \quad (19)$$

and Eq. (18) becomes

$$\ddot{\Phi}_n = \frac{k}{m}(\Phi_{n+1} + \Phi_{n-1} - 2\Phi_n) - \frac{K}{m}(\Phi_{n+h} + \Phi_{n-h} + 2\Phi_n) - \omega_g^2(\Phi_n + \varepsilon \alpha \Phi_n^2 + \varepsilon^2 \beta \Phi_n^3), \quad (20)$$

where

$$\omega_g^2 = \frac{4a^2 D}{m}, \quad \alpha = \frac{-3a}{\sqrt{2}} \quad \text{and} \quad \beta = \frac{7a^2}{3}. \quad (21)$$

Now, we solve Eq. (20) following [7], where all derivations can be found. To solve it, we use a semi-discrete approximation, which means that we look for the wave solutions of the form

$$\Phi_n(t) = F_1(\xi)e^{i\theta_n} + \varepsilon[F_0(\xi) + F_2(\xi)e^{i2\theta_n}] + cc + O(\varepsilon^2) \qquad (22)$$

$$\xi = (\varepsilon nl, \varepsilon t), \qquad \theta_n = nql - \omega t, \qquad (23)$$

where $l = 0.34$nm is the distance between two neighboring nucleotides in the same strand, $\omega$ is the optical frequency of the linear approximation, $q = 2\pi/\lambda$ is the wavenumber, cc represents complex conjugate terms and the function $F_0$ is real. This is a modulated wave where $F_1$ is a continuous function representing the envelope while the carrier component $e^{i\theta_n}$ is discrete. As the frequency of the carrier wave is much higher than the frequency of the envelope we need the two time scales, $t$ and $\varepsilon t$, for those two functions, which can be seen in Eq. (23). Of course, the same holds for the coordinate scales. A mathematical basis for this procedure is a multiple-scale method or a derivative-expansion method [35,36].

From Eq. (22), one can see the true meaning of the parameter $\varepsilon$. The higher-order terms are required because of the last two terms in Eq. (20).

Now, we switch to the continuum limit $nl \to z$ and use the transformations

$$Z = \varepsilon z; \qquad T = \varepsilon t, \qquad (24)$$

which yield to

$$\Phi_n(t) \to F_1(Z,T)e^{i\theta} + \varepsilon[F_0(Z,T) + F_2(Z,T)e^{i2\theta}] + cc$$
$$= F_1 e^{i\theta} + \varepsilon[F_0 + F_2 e^{i2\theta}] + F_1^* e^{-i\theta} + \varepsilon F_2^* e^{-i2\theta}, \qquad (25)$$

where the star stands for complex conjugate and $F_i \equiv F_i(Z,T)$. Also, very important relation is

$$F_i(\varepsilon(n \pm h)l, \varepsilon t) \to F_i(Z,T) \pm F_{iZ}(Z,T)\varepsilon lh + \frac{1}{2}F_{iZZ}(Z,T)\varepsilon^2 l^2 h^2, \qquad (26)$$

where index $Z$ denotes differentiation [7]. All this enables us to determine the expressions existing in Eq. (20), such as

$$\Phi_{n+h} + \Phi_{n-h} + 2\Phi_n = \{2F_1[\cos(qhl)+1] + 2i\varepsilon hlF_{1Z}\sin(qhl) + \varepsilon^2 h^2 l^2 F_{1ZZ}\cos(qhl)\}e^{i\theta}$$
$$+ \{2\varepsilon F_2[\cos(2qhl)+1] + 2i\varepsilon^2 hlF_{2Z}\sin(2qhl)\}e^{i2\theta} + 4\varepsilon F_0 + cc, \qquad (27)$$

$$\dot{\Phi}_n = \varepsilon F_{1T}e^{i\theta} - i\omega F_1 e^{i\theta} + \varepsilon^2 F_{0T} + \varepsilon^2 F_{2T}e^{i2\theta} - 2i\varepsilon\omega F_2 e^{i2\theta} + cc \qquad (28)$$

and

$$\Phi_n^3 = 3|F_1|^2 F_1 e^{i\theta} + 3|F_1|^2 F_1^* e^{-i\theta} + F_1^3 e^{i3\theta} + F_1^{*3} e^{-i3\theta} + O(\varepsilon). \qquad (29)$$

Finally, we obtain the continuum version of Eq. (20), that is

$$(\varepsilon^2 F_{1TT} - 2i\varepsilon\omega F_{1T} - \omega^2 F_1)e^{i\theta} - (4i\varepsilon^2\omega F_{2T} + 4\varepsilon\omega^2 F_2)e^{i2\theta} + cc =$$
$$= \frac{k}{m}\{2F_1[\cos(ql)-1]e^{i\theta} + 2i\varepsilon lF_{1Z}\sin(ql)e^{i\theta} + \varepsilon^2 l^2 F_{1ZZ}\cos(ql)e^{i\theta}$$
$$+ 2\varepsilon F_2[\cos(2ql)-1]e^{i2\theta} + 2i\varepsilon^2 lF_{2Z}\sin(2ql)e^{i2\theta} + cc\}$$
$$- \frac{K}{m}\{2F_1[\cos(qhl)+1]e^{i\theta} + 2i\varepsilon hlF_{1Z}\sin(qhl)e^{i\theta} + \varepsilon^2 h^2 l^2 F_{1ZZ}\cos(qhl)e^{i\theta}$$
$$+ 2\varepsilon F_2[\cos(2qhl)+1]e^{i2\theta} + 2i\varepsilon^2 hlF_{2Z}\sin(2qhl)e^{i2\theta} + 4\varepsilon F_0 + cc\}$$
$$- \omega_g^2\Big[F_1 e^{i\theta} + \varepsilon F_0 + \varepsilon F_2 e^{i2\theta} + 2\varepsilon\alpha|F_1|^2 + 2\varepsilon^2\alpha(F_0 F_1 + F_1^* F_2)e^{i\theta}$$
$$+ \varepsilon\alpha F_1^2 e^{i2\theta} + 2\varepsilon^2\alpha F_1 F_2 e^{i3\theta} + 3\varepsilon^2\beta|F_1|^2 F_1 e^{i\theta} + \varepsilon^2\beta F_1^3 e^{i3\theta} + cc\Big]. \quad (30)$$

Let us keep in mind that we are looking for the solution $\Phi_n(t)$. The crucial expression (30) enables us to obtain the functions $F_1(\xi)$, $F_0(\xi)$ and $F_2(\xi)$, required for the determination of $\Phi_n(t)$, as clear from Eq. (22). To do this we should equate the coefficients for the various harmonics [6,7,33]. Thus, the coefficients for $e^{i\theta}$ give

$$\varepsilon^2 F_{1TT} - 2i\varepsilon\omega F_{1T} - \omega^2 F_1 = \frac{k}{m}\Big[2F_1(\cos(ql)-1) + 2i\varepsilon lF_{1Z}\sin(ql) + \varepsilon^2 l^2 F_{1ZZ}\cos(ql)\Big]$$
$$- \frac{K}{m}\Big[2F_1(\cos(qhl)+1) + 2i\varepsilon hlF_{1Z}\sin(qhl) + \varepsilon^2 h^2 l^2 F_{1ZZ}\cos(qhl)\Big]$$
$$- \omega_g^2\Big[F_1 + 2\varepsilon^2\alpha F_0 F_1 + 2\varepsilon^2\alpha F_1^* F_2 + 3\varepsilon^2\beta|F_1|^2 F_1\Big]. \quad (31)$$

Neglecting all the terms with $\varepsilon$ and $\varepsilon^2$ we get the dispersion relation

$$\omega^2 \equiv \omega_y^2 \equiv \omega_o^2 = (4/m)\big[a^2 D + k\sin^2(ql/2) + K\cos^2(qhl/2)\big], \quad (32)$$

which brings about the expression for the group velocity $d\omega/dq$ as

$$V_g = \frac{l}{m\omega}[k\sin(ql) - Kh\sin(qhl)]. \quad (33)$$

The corresponding dispersion relation for the in-phase oscillations described by (17) is

$$\omega_x^2 \equiv \omega_a^2 = (4/m)\big[k\sin^2(ql/2) + K\sin^2(qhl/2)\big]. \quad (34)$$

The frequencies $\omega_y$ and $\omega_x$ are usually called optical and acoustical.

Equating the coefficients for $e^{i0} = 1$, we straightforwardly obtain

$$F_0 = \mu|F_1|^2; \qquad \mu = -2\alpha\left(1 + \frac{4K}{m\omega_g^2}\right)^{-1}, \quad (35)$$

while $e^{i2\theta}$ gives

$$F_2 = \delta F_1^2; \qquad \delta = \omega_g^2 \, \alpha \left[ 4\omega^2 - \frac{4k}{m} \sin^2(ql)) - \frac{4K}{m} \cos^2(2hql) - \omega_g^2 \right]^{-1}. \qquad (36)$$

As the functions $F_0$ and $F_2$ can be expressed through $F_1$ the equation for $F_1$ should be derived. We use the new coordinates

$$S = Z - V_g T, \qquad \tau = \varepsilon T, \qquad (37)$$

again and obtain the transformations for $F_{1Z}$, $F_{1ZZ}$, $F_{1T}$ and $F_{1TT}$ existing in Eq. (30). We notice that $\varepsilon$ exists in the time coordinate but does not in the space one. This definition ensures that the time variation of the envelope of the function $F_1$, in units of $1/\omega$, be smaller than its space variation in units of $l$ [37]. Finally, using Eqs. (31)-(33) and (35)-(37), we easily obtain the well-known nonlinear Schrödinger equation (NLSE) for the function $F_1$

$$iF_{1\tau} + P F_{1SS} + Q \, |F_1|^2 F_1 = 0, \qquad (38)$$

where the dispersion coefficient $P$ and the coefficient of nonlinearity $Q$ are given by

$$P = \frac{1}{2\omega} \left\{ \frac{l^2}{m} [k \cos(ql) - Kh^2 \cos(qhl)] - V_g^2 \right\}, \qquad Q = -\frac{\omega_g^2}{2\omega} [2\alpha(\mu + \delta) + 3\beta]. \qquad (39)$$

This is a solvable equation and its analytical solution, for $PQ > 0$, is [33,38,39]

$$F_1(S,\tau) = A_0 \, \text{sech}\left(\frac{S - u_e \tau}{L_e}\right) \exp \frac{iu_e(S - u_c \tau)}{2P}, \qquad u_e > 2u_c. \qquad (40)$$

Here, we assume $P > 0$ and $Q > 0$ [38]. The values for the envelope amplitude $A_0$ and its width $L_e$ will be written later. The function (40) is obviously the modulated solitary wave, where $u_e$ and $u_c$ are the velocities of the envelope and the carrier waves, respectively.

Therefore, we have obtained the expression for $F_1$, the functions $F_0$ and $F_2$ can be expressed through $F_1$ and, according to Eqs. (19), (22) and (23), and we can easily reach our final goal that is the function $y_n(t)$.

However, before we proceed, we need to comment a couple of the parameters existing in the HPB model. Some of them have appeared in the Hamiltonian (16). They are so-called intrinsic parameters, describing the geometry and the chemical interactions within DNA. However, there are the parameters coming from the applied mathematical procedure. Let us concentrate on the mathematical parameters $u_e$, $u_c$ and $\varepsilon$. The velocities $u_e$ and $u_c$ are included in the solution of the NLSE, while $\varepsilon$ does not have any physical meaning. This only helps us to distinguish big and small terms in the series expansion (22). A careful investigation of all the formulae shows that only two mathematical parameters are relevant and they are: $\varepsilon u_e$ and $\varepsilon u_c$. Also, it is very difficult to pick the appropriate values for $u_e$ and $u_c$ according to the requirement $u_e > 2u_c$ only. However, the ratio of these speeds belongs to the interval [0;0.5), which is much more convenient to deal with. Hence, we choose the following two mathematical parameters [40]

$$U_e = \varepsilon u_e, \qquad \eta = \frac{u_c}{u_e}, \qquad 0 \leq \eta < 0.5. \qquad (41)$$

We will return to this point later and show how $U_e$ can be expressed through $\eta$.

Finally, according to the expressions (19), (22)-(24), (35), (36), (38) and (41), the stretching of the nucleotide pair at the position $n$, i.e., the solution of Eq. (18), is

$$y_n(t) = 2A \operatorname{sech}\left(\frac{nl - V_e t}{L}\right) \left\{ \cos(\Theta nl - \Omega t) + A \operatorname{sech}\left(\frac{nl - V_e t}{L}\right) \left[\frac{\mu}{2} + \delta \cos(2(\Theta nl - \Omega t))\right] \right\}, \qquad (42)$$

where

$$A \equiv \varepsilon A_0 = |U_e| \sqrt{\frac{1 - 2\eta}{2PQ}}, \qquad L \equiv \frac{L_e}{\varepsilon} = \frac{2P}{|U_e|\sqrt{1 - 2\eta}}. \qquad (43)$$

The envelope velocity $V_e$, the wavenumber $\Theta$ and the frequency $\Omega$ are given by

$$V_e = V_g + U_e, \qquad \Theta = q + \frac{U_e}{2P}, \qquad \Omega = \omega + \frac{(V_g + \eta U_e)U_e}{2P}. \qquad (44)$$

To plot the function $y_n(t)$ the values of all the parameters should be known. The problem with the mathematical parameter $U_e$ can be solved assuming that the most favorable mode is a coherent one (CM). This assumes that the envelope and the carrier wave velocities are equal, i.e., [41]

$$V_e = \frac{\Omega}{\Theta}. \qquad (45)$$

This means that the function $y_n(t)$ is the same at any position n. In other words, the wave preserves its shape in time, indicating high stability [40]. Notice that the requirement (45) ensures that Eq. (42) becomes one phase function. This means that $y_n(t)$ depends on $nl$ and $t$ through $\xi = nl - V_e t$, where $V_e$ is a constant, representing the travelling wave. If a solitary wave, or soliton for short, is defined as a localized travelling wave [39], then $y_n(t)$, obviously being localized, satisfies the requirements for being the soliton. In other words, the CM is nothing but the solitonic mode (SM) [7]. According to Eq. (45), one can easily obtain the function $U_e(\eta)$, which is

$$U_e = \frac{P}{1-\eta}\left[-q + q\sqrt{1 + \frac{2(1-\eta)}{Pq^2}(\omega - qV_g)}\right]. \tag{46}$$

This is a slowly increasing function of $\eta$ [40].

Therefore, we have solved the problem regarding the parameter $U_e$. Let us study $\eta$ now. It was pointed out that Eq. (42) represents the modulated wave. It is useful to define a certain physical quantity, which determines the efficiency of the modulation. This can be a density of internal oscillations (density of carrier wave oscillations) [41]. This is a ratio of wavenumbers of the wave components, or a ratio of the appropriate periods. According to Eq. (42), we can define the wavelengths and periods of both the envelope and carrier wave as

$$\frac{2\pi}{\Lambda} = \frac{1}{L}, \quad \frac{2\pi}{\Lambda_c} = \Theta, \quad \frac{2\pi}{T} = \frac{V_e}{L}, \quad \frac{2\pi}{T_c} = \Omega, \tag{47}$$

where the index $c$ denotes the carrier component. Then, we define the density in two ways, that is

$$D_o \equiv \frac{\Lambda}{\Lambda_c} = L\Theta, \qquad \Gamma_o \equiv \frac{T}{T_c} = \frac{L\Omega}{V_e}. \tag{48}$$

They are equal as should be if Eq. (45) is satisfied. This is strong support to the CM. In what follows, we plot $D_o(\eta)$, as well as $A_m(\eta) = 2A$, and try to pick an acceptable value for the parameter $\eta$. However, to do these plots, we should know, or assume, the values of the remaining intrinsic parameters. They are: $k$, $K$, $a$ and $D$, existing in Eq. (16), and $q$, which has appeared in Eq. (23). The experimental values of these parameters do not exist and this has turned out to be a very tough problem. It is very likely that the most detailed analysis has been performed in Ref. [42]. It was shown that there are a couple of requirements that should be satisfied. In this chapter, we use the following set of the parameters, satisfying all these requirements [7,42]:

$$a = 1.2\,\text{Å}^{-1}, \quad D = 0.07\,\text{eV}, \quad k = 12\,\text{N/m}, \quad K = 0.08\,\text{N/m}, \quad ql = 2\pi/10. \tag{49}$$

Now, we can plot the functions $D_o(\eta)$ and $A_m(\eta)$. They are shown in Figs. 4 and 5, respectively. The plots were done according to Eqs. (21), (32), (33), (35), (36), (39), (43), (44), (46), (48) and (49). We see that for very small $\eta$ modulation, practically, does not exist. If we assume that modulation represents a key factor in DNA dynamics then it is very likely that $\Gamma_o$ cannot be less than 6. This means that $\eta$ should be around 0.45 or bigger. Figure 5 also shows that the small values for $\eta$ are not acceptable. If we expect that $A_m$ cannot be bigger than 0.3 we come up with the same conclusion as above. Notice that $A_m = 2A$ is not real amplitude. From Eq. (42) we see that $y_n(t)$ has maximal value when the cosine and secant hyperbolic functions are equal to one. Such a curve is similar to one depicted in Fig. 5 but is roughly 1.5 times bigger. For example, the starting point for $\eta = 0$ would be 0.97 instead of 0.63 shown in Fig. 5. As a conclusion, in what follows, we use $\eta = 0.47$.

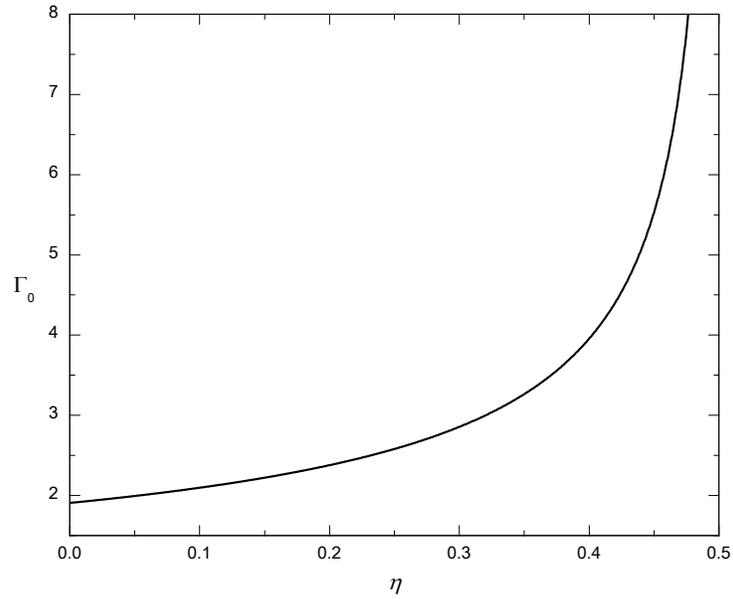

Figure 4: Density of internal oscillations $\Gamma_o$ as function of the parameter $\eta$

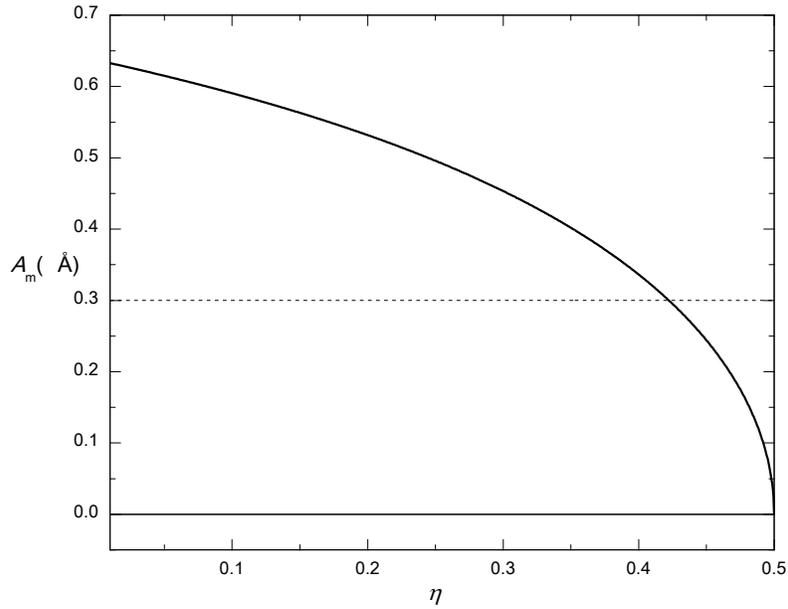

Figure 5: Amplitude $A_m = 2A$ as function of the parameter $\eta$

Finally, we can plot the nucleotide pair stretching as a function of the position, i.e., the function $y_n(t)$. This is depicted in Fig. 6 as a function of the position for a particular time $t$.

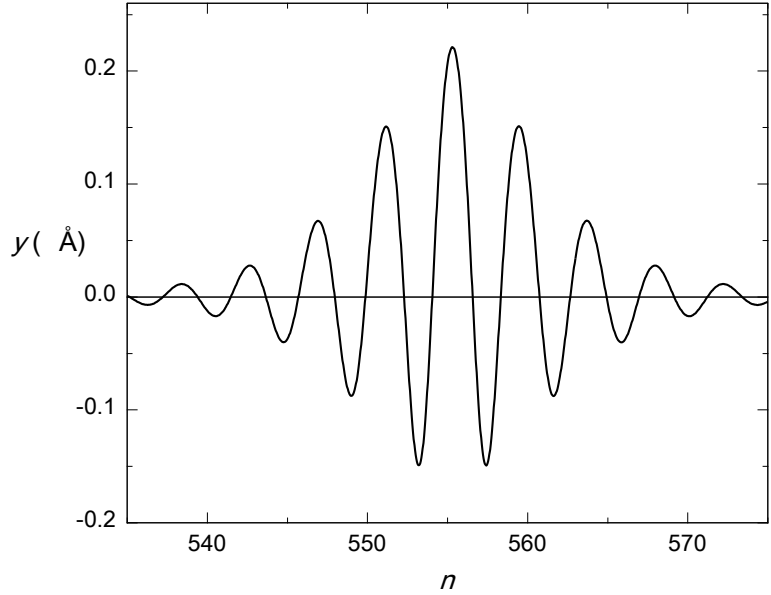

Figure 6: Nucleotide pair stretching at $t = 100\text{ps}$ for: $a = 1.2\,\text{Å}^{-1}$, $D = 0.07\,\text{eV}$, $k = 12\,\text{N/m}$, $K = 0.08\,\text{N/m}$, $ql = 2\pi/10$ and $\eta = 0.47$.

It is obvious that this is a localised modulated wave, usually called breather. If we had picked a different time we would have obtained exactly the same shape of the wave but at a different position. This is so because the CM was assumed. Also, we could have assumed a certain position and plot $y_n(t)$ as a function of time.

According to Fig. 6, one can see that the positive amplitude is a little bit bigger than the negative one. This is coming from the higher-order term in (22). Basically, this is a result of the fact that the Morse potential is not symmetric, which means that the repulsive force between the nucleotides is stronger than the attractive one.

Based on Fig. 6, we can conclude that the soliton covers about 30 nucleotide pairs. In other words, the wavelength, defined by Eq. (47), is $\Lambda = 30.2l$. Unfortunately, appropriate experimental values do not exist. However, this width can be compared with the solitonic width at a DNA segment involved in a process of transcription [7]. It was reported [43] that this width is between 8 and 17 nucleotides, while some experimental works suggest that this segment covers between 7 and 15 base pairs [44]. The width shown in Fig. 6 is higher, but we should keep in mind that the transcription is followed by a local unzipping, which can be understood as an extremely high amplitude. The wave shown in Fig. 6 is an "ordinary" one, while the solitons at the mentioned segments have much higher amplitudes, corresponding to the local unzipping, which is a topic of the next sections. As the increase of the amplitude means the decrease of the solitonic width, we can conclude that the solitonic width, corresponding to Fig. 6, certainly makes sense.

It was stated above that the two basic improvements of the PB model have been done so far. We have just described the HPB model in some more details. Now, we briefly explain the PBD model. The basic idea is that the harmonic potential energy has been replaced by the anharmonic one through

$$\frac{k}{2}(y_n - y_{n-1})^2 \quad \Rightarrow \quad \frac{k}{2}\left[1 + \rho e^{-\alpha(y_n + y_{n-1})}\right](y_n - y_{n-1})^2, \tag{50}$$

where $\rho$ and $\alpha$ are constants [45-50]. This expression can be viewed as a harmonic interaction with a variable coupling constant [51].

It was stated above that $y_n(t)$ is not temperature-dependent but its mean value is. Figure 7, reproduced from Ref. [46], shows how the mean value $\langle y \rangle$ depends on temperature. The authors compared the two cases within the potential given by Eq. (50), that is $\alpha = 0$ (PB model) and $\alpha \neq 0$ (PBD model). We see that $\langle y \rangle$ is slowly increasing function up to a certain temperature when it sharply increases. This increase represents denaturation. In the case of the PBD model, denaturation is rather sharp and occurs at lower temperatures. It would be interesting to study temperature dependences of $\langle y \rangle$ relying on the HPB, as well as other models, which could be a future task. We want to point out that temperature dependence of $\langle y \rangle$ also depends on the remaining parameters existing in the model, like $D$ and $a$, describing the Morse potential.

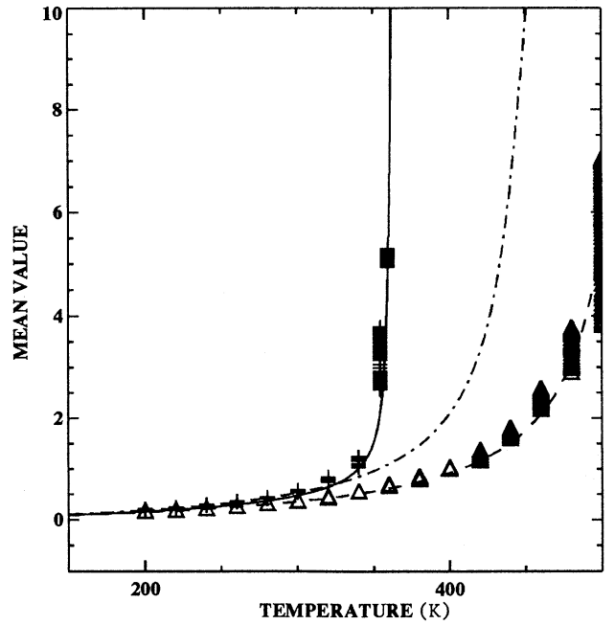

Figure 7: Variation of the mean value $\langle y \rangle$ vs. temperature for: $k = 0.04\text{eV}/\text{Å}^2$, $a = 4.45\,\text{Å}^{-1}$, $D = 0.04\text{eV}$. The solid line corresponds to the anharmonic stacking interaction ($\alpha = 0.35\,\text{Å}^{-1}$, $\rho = 0.5$). The dashed and dash-dotted lines correspond to two cases ($\rho = 0.5$ and $\rho = 0$) of harmonic stacking interactions ($\alpha = 0$), respectively.

The models explained above assume homogeneous DNA chain, which might not be quite correct. For example, Adenine and Thymine are connected by double and Guanine and Cytosine by triple hydrogen bonds. Hence, one can expect that the corresponding Morse potential depths are $D$ and $1.5D$. A crucial question is whether the wave characteristics like amplitude, velocity, etc., drastically change whenever the wave reaches a new type of the nucleotide pair. If so, the soliton would not be stable. It was explained that the breathers are not substantially affected by spatial inhomogeneities of the DNA sequence, but the kinks are [18]. This is in a good agreement with the result explained in [52].

## 2. Resonance mode and DNA opening

Any model can be considered as good if it can explain something or predict a possible experiment. In this section, we show how the HPB model can explain a local opening of DNA, the well-known fact which happens during transcription. Let us study the functions $\omega_o(ql)$ and $\omega_a(ql)$, given by Eqs. (32) and (34), respectively. They also depend on the parameters $k$, $K$, $a$ and $D$, and, in general, are not equal. These frequencies were compared [33,53] and it was speculated that their equality could represent a resonance mode (RM) [53]. Let us study this idea in some more details. From Eqs. (32) and (34) one obtains

$$\omega_o^2 - \omega_a^2 \propto \frac{a^2 D}{K} + \cos(5ql). \tag{51}$$

We easily recognise the following three possibilities:

1) $a^2 D > K \quad \Rightarrow \quad \omega_o > \omega_a \quad \text{for} \quad \forall ql,$ \hfill (52a)

2) $a^2 D < K \quad \Rightarrow \quad \omega_o < \omega_a \quad \text{in some intervals of } ql,$ \hfill (52b)

3) $a^2 D = K \quad \Rightarrow \quad \omega_o = \omega_a \quad \text{at } ql = \pi/5.$ \hfill (52c)

Of course, there are other values for $ql$ satisfying the requirement $\omega_o = \omega_a$, but they are not relevant now. The last case, i.e., Eq. (52c), corresponds to the RM, mentioned above [53]. The idea was further developed in [54,55]. It was shown that some conditions yield to very high amplitude [54]. Of course, the large amplitude does not necessarily mean the RM and this behavior was called extremely high amplitude (EHA) mode in [54]. However, some arguments in favor of the RM were given in Ref. [55] and, in this chapter, the term RM will be used.

    Before we proceed, we want to discuss one important point. Equations (17) and (18) represent two decoupled equations of motion. How about the corresponding frequencies, given by Eqs. (32) and (34)? We should notice that $\omega_o$ and $\omega_a$ are not decoupled in a sense that they can be changed independently as both frequencies depend on the same parameters $k$ and $K$. Hence,

they are coupled through the common parameters [54]. In other words, we can eliminate one of these parameters and express $\omega_o$ as a function of $\omega_a$, or vice versa.

In what follows, we explain how the RM can explain the local opening of the DNA chain during the mRNA formation. The DNA-RNA transcription is nothing but the formation of mRNA molecule from RNA polymerase molecules (RNAP), as shown in Fig. 8 [5,56]. Therefore, this occurs at the segments where the DNA chain is surrounded by the RNAP. We can call them transcription segments (TS). The transcription can be done only through an active interaction between DNA and its surrounding. Let us imagine a segment of one DNA strand. The nucleotides belonging to this strand can interact with the surrounding nucleotides only if they do not interact strongly with the remaining strand, which can be seen in Fig. 8. This means that DNA chain should open locally and this is what really happens during the transcription.

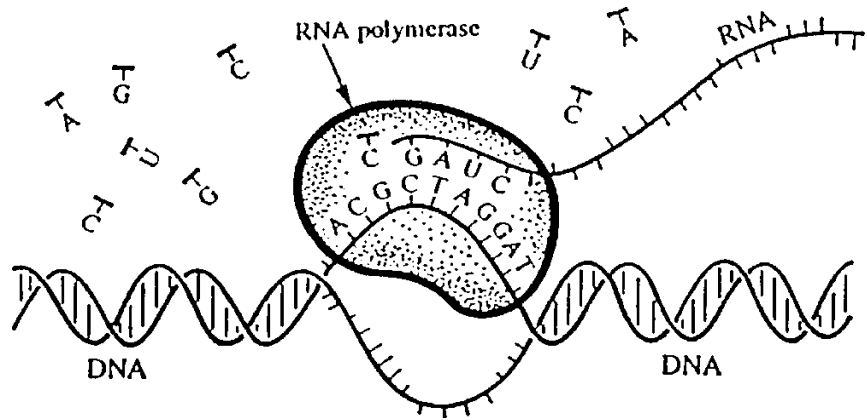

Figure 8: Transcription of RNA

It was explained above that the mentioned parameters describe chemical bonds. This means that, due to the presence of RNAP, the interaction between the nucleotides belonging to the same pair is changed. In other words, the Morse potential at TSs is different from the potential at the rest of the molecule, which means that the RNAP transforms the Morse potential so that it becomes wider and shallower. This change could correspond to the transition from the blue to red cases in Fig. 2. Notice that the result of the local opening is the decrease in the parameter $K$. Therefore, the RNAP interacts with the DNA nucleotides, decreasing the force between the different strands. This means that both $K$ and $a^2D$ go down, $a^2D$ decreasing faster, and, finally, the EHA vibrations occur ($K = a^2D$), which results in very big amplitude i.e., in the local opening of the DNA chain.

It is worth mentioning that the decrease of the value of the parameter $K$ might be an indication that the helicoidal interaction term in Eq. (16) is not linear. In other words, it might make sense to replace this term with the nonlinear one. This has not been done so far but might be one of the future tasks.

One can argue that $K = a^2D$ is not a sufficient condition for EHA vibrations, i.e., for the RM, as $ql = \pi/5$ is required in addition. However, it was explained above that $ql = \pi/5$ is probably the most favorable value, as indicated in Eq. (49). This is nothing but $N = 10$ in Fig. 6.

The reader might have noticed that the optical frequency is $\Omega$, given by Eq. (44), rather than $\omega_o$. This is correct but the simplified analysis has been quite appropriate till now to understand the idea. Therefore, a more correct requirement for the RM is

$$\Omega \equiv \omega_o + \Delta\omega = \omega_a, \tag{53}$$

while $\omega_o = \omega_a$ represents its approximation. Of course, the expression for $\Delta\omega$ is determined by Eq. (44).

Till now, we have discussed the local opening and RM, but there was nothing that can be considered as a possible proof for the real existence of the RM. The best and the easiest that can be done is to study the amplitude, existing in Eq. (42). We should investigate if it really goes to infinity under the mentioned conditions. It is convenient to introduce a positive dimensionless parameter $p$ defined through

$$a = p\sqrt{K/D}. \tag{54}$$

Obviously, the cases $p > 1$, $p < 1$ and $p = 1$ correspond to Eqs. (52a-c), respectively.

Figure 9 shows the amplitude as a function of the parameter $p$ for three values of $D$. It was plotted according to Eqs. (21), (32), (33), (35), (36), (39), (43) and (46). Of course, the chosen values for $D$ and $K$ are smaller than the non-resonant ones, used for Fig. 6. We can see that the amplitude really tends to infinity for a certain critical value of $p$. This is smaller than one due to the definition (53).

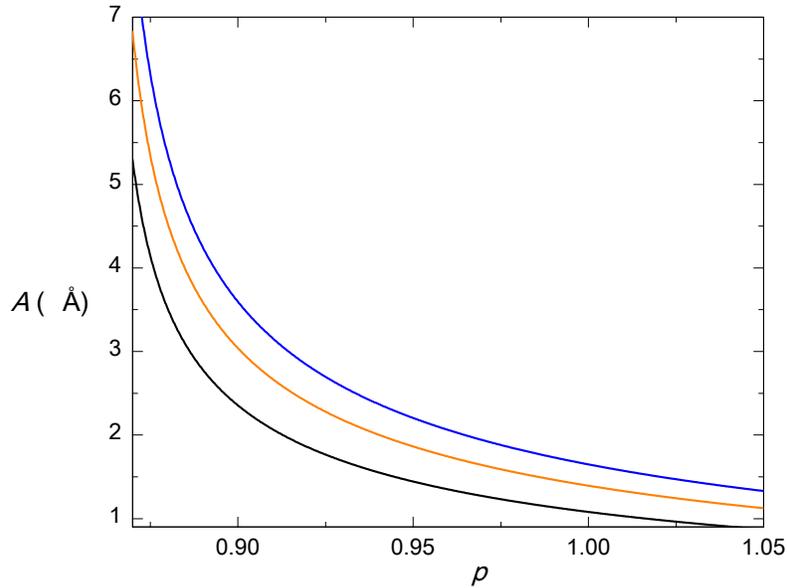

Figure 9: Amplitude $A$ as a function of $p$ for $k = 12\,\text{N/m}$, $K = 0.05\,\text{N/m}$ and $D = 0.07\,\text{eV}$ (blue), $D = 0.05\,\text{eV}$ (red) and $D = 0.03\,\text{eV}$ (black)

Two examples of the resonance solitons $y(n)$ are shown in Figs. 10 and 11. These functions were plotted like Fig. 6, but $a$ was determined according to Eq. (54). It was explained above that the RM values of the parameters $K$ and $a^2 D$ are smaller than the ordinary values and we picked $K = 0.05\,\text{N/m}$ and $D = 0.05\,\text{eV}$. Both figures were carried out for $t = 100\,\text{ps}$. It is important to point out that the interaction between the RNAP and the corresponding DNA nucleotides does not affect the longitudinal interaction of the neighboring nucleotides. Hence, the parameter $k$ is not changed at those segments unlike $K$, $a$ and $D$. Notice that $k$ does not appear in Eq. (54). Of course, very high amplitude should not bother us as viscosity has been neglected.

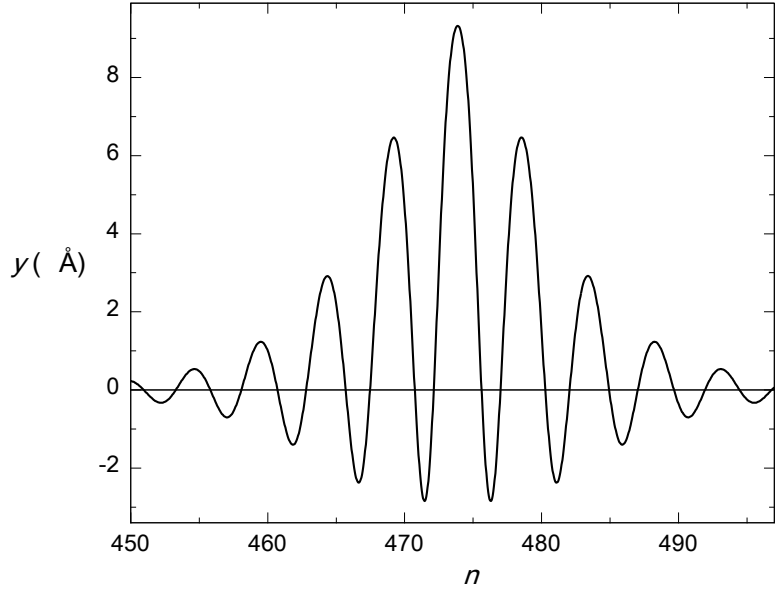

Figure 10: Nucleotide pair stretching at $t = 100\,\text{ps}$ for: $D = 0.05\,\text{eV}$, $k = 12\,\text{N/m}$, $K = 0.05\,\text{N/m}$, $ql = 2\pi/10$, $\eta = 0.47$ and $p = 0.9$

It was explained why the positive amplitude in Fig. 6 is slightly bigger than the negative one. However, for the RM, the negative amplitude becomes negligible in comparison to the positive one. This is in the agreement with our attempt to describe the local opening of the DNA molecule.

It is very important to understand that the resonance cannot happen for $K = 0$. This means that the helicoidal structure provides the resonance, which shows the advantage of the HPB model over the PB one, which could be obtained from the HBD by letting $K = 0$. There is one more way to demonstrate this advantage. The figure 4 shows how modulation, i.e., the density of internal oscillations $\Gamma_o$, depends on $\eta$ for the particular $K$. We could have assumed a fixed value for $\eta$ and plotted the function $\Gamma_o(K)$. This is an increasing function showing that the higher $K$ provides more efficient modulation.

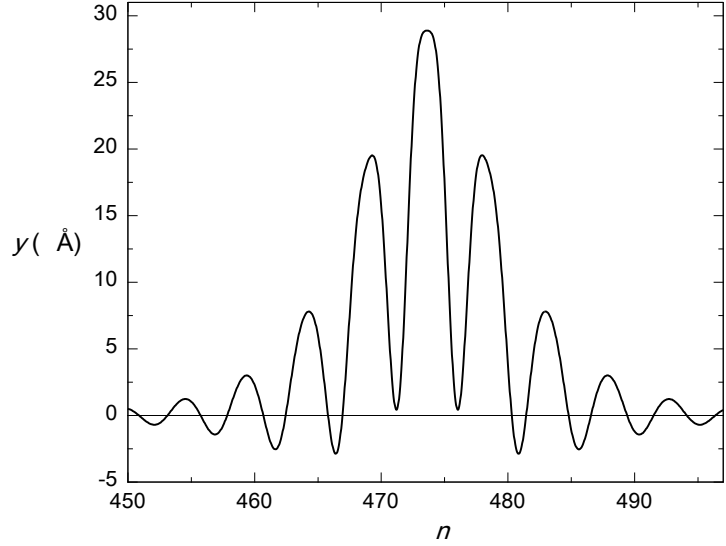

Figure 11: Nucleotide pair stretching at $t = 100\,\text{ps}$ for: $D = 0.05\,\text{eV}$, $k = 12\,\text{N/m}$, $K = 0.05\,\text{N/m}$, $ql = 2\pi/10$, $\eta = 0.47$ and $p = 0.87$

Let us study the soliton depicted in Fig. 11. According to Eq. (47), its width, in units of $l$, is $\Lambda_n \equiv \Lambda/l \approx 38$. However, the local opening can be related to its positive value only. We see that such relevant width is about 13 nucleotides. This can be compared with the experimental values, where the transcription bubble and RNA:DNA hybrid were reported to be 17 and 8 base pairs, respectively [57]. In [54], where slightly different values of the parameters were used, the agreement with these experimental values was almost perfect. However, we should not be too happy with providing such agreement because, more or less, everything depends on a few parameters and, of course, different combinations of them can bring about an equal result.

More detailed analysis of the RM exists in [54]. There, it was shown how the interval of the allowed values of $K$ can be estimated. Also, the smallest value of $p$ was calculated. Of course, this is the resonance value $p_R$. For the values of the parameters used in this paragraph, $p_R$ is slightly smaller than 0.87, used for Fig. 11.

Figure 9 shows that the amplitude reaches extremely high values under certain conditions. However, each big amplitude does not necessarily mean RM. In the context of classical mechanics, the vibrations of an undamped linear oscillator, characterized by intrinsic frequency $\omega_i$, when subjected under the action of external harmonic force of frequency $\omega_f$, must attain a resonance regime when $\omega_i = \omega_f$. How about the DNA molecule? Let us imagine an arbitrary nucleotide pair. If this pair were independent, i.e., free from any external influence, the center of its mass would not move. This means that there would be only one mode and these are out-of-phase oscillations. Therefore, the frequency $\omega_o$, given by Eq. (32), corresponds to the intrinsic frequency $\omega_i$ of the classical undamped oscillator. However, when this nucleotide pair belongs to DNA there is one more oscillating mode and the frequency $\omega_a$. In other words, there is surrounding, which brings about one more mode. From the point of view of the nucleotide pair this surrounding, i.e., the rest of DNA, is nothing but the external force [55]. Hence, we can expect the resonance mode to happen if these frequencies are equal as the friction is neglected and this is what we have stated in the paper. Therefore, the frequencies $\omega_o$ and $\omega_a$ correspond to the frequencies $\omega_i$ and $\omega_f$, respectively [55].

At the beginning of this chapter, a couple of very important and relevant years were mentioned. Now, let us remember one more. This is 1992 when the first mechanical manipulation on a single molecule was performed [58]. Therefore, it is possible to stretch, wind, unwind and so on, the single molecule [59-70]. The first molecule that was picked for such experiments was DNA [58]. There have been some suggestions for the experiments that would test the theory explained in this section [71]. It is very likely that such experiments are still not realistic to be performed. However, the value of the parameter $k$ can be determined, as was suggested recently [72]. The idea for the experiment is based on the fact that the longitudinally applied force on DNA is proportional, in a rather big interval, to its extension [72]. It is important to point out that this linearity represents proof that the longitudinal interaction along DNA (the second term in Eq. (13)), is properly modeled by the harmonic potential. Of course, if we knew the value of $k$ we would be able to estimate the remaining parameters better.

It was mentioned above that the very high amplitude should not bother us because viscosity has been neglected so far. Otherwise, the infinitely large-amplitude would represent the destruction of the molecule. To provide a more realistic DNA model, we should take viscosity into consideration. This can be done by adding a viscous force

$$F_v = -\gamma \, \dot{y}_n \tag{55}$$

into Eq. (18), where $\gamma$ represents a damping coefficient [73-75]. The optical frequency and the group velocity become [7,75]

$$\omega_\gamma + i\chi = \sqrt{\omega^2 - \chi^2}, \qquad V_\gamma = \frac{\omega V_g}{\sqrt{\omega^2 - \chi^2}}, \qquad \chi = \gamma/2m. \tag{56}$$

We can follow the same procedure as above and this brings about Eq. (38) again, but the expressions for the dispersion and nonlinear parameters are different. A crucial point is a fact that the complex optical frequency yields to the complex $Q$ and Eq. (38) cannot be solved analytically any more. A numerical solution is very interesting from the biological point of view [75]. The obtained wave looks like the envelope of the soliton in Fig. 6 with smaller amplitude. This means that viscosity destroys modulation and the wave is bell-type soliton. This has very important biological implication, which will be explained later. It suffices now to state that the demodulation ensures long-lasting interaction between DNA nucleotides and RNAP, which is biologically very convenient [75].

## 3. Demodulated standing solitary wave and DNA-RNA transcription

In this subsection, we study DNA:RNA transcription again. Hence, we keep in mind a certain TS and Fig. 8. However, instead of the RM, we study the transcription in the context of two new ideas. We rely on the HPB model again and follow [72].

Let us concentrate on a particular DNA nucleotide in Fig. 8. If this is an adenine, for example, it is bonded with DNA thymine, belonging to other strand, but also interacts with RNAP. Of course, the final positioning of RNA nucleotides should be a copy of the DNA segment. This obviously means that our DNA adenine should attract a certain RNA uracile and repel the remaining RNA nucleotides. This can be efficiently done only if the DNA adenine is far enough from its DNA partner during the transcription. Of course, this is really the case due to the local opening, as explained above.

Now, we go further in this direction of thinking. The local opening is certainly necessary but not sufficient condition for successful transcription. The stretching of DNA, i.e., the distance between the DNA nucleotides belonging to the same pair, is described by Eq. (42). This certainly means that our adenine and thymine are far from each other during short periods of time only and the adenine we have in mind does not have enough time to attract one RNA uracile. We do believe that the carrier wave is crucial for soliton movement through DNA chain but is redundant when transcription occurs. Also, it is clear that only the envelope of Eq. (42) may correspond to the local opening. All this suggests the idea that the breather, moving along the chain, should be demodulated when it reaches the TSs. Mathematical interpretation of this requirement is

$$\Theta = 0, \qquad \Omega = 0, \qquad (57)$$

as can be concluded according to Eq. (42). A crucial question is how demodulation happens at these segments. Our explanation is, like above, that RNAP changes chemical milieu of DNA nucleotides, i.e., the values of relevant parameters, especially $D$ and $a$, which yields to the values accommodating Eq. (57).

Therefore, the lack of internal oscillations prolongs the interaction between our adenine and thymine, but the question is if this is enough for successful transcription. We should keep in mind that RNAP, during transcription, normally processes up to 100 bps per second [5]. This corresponds to propagation velocities of $34\,\text{nm/s}$, which is negligibly small in comparison to soliton velocities in DNA. Hence, a biologically convenient soliton is the one that is as slow as possible at TSs as this would decrease the probability for the genetic mistakes as much as possible. If we assume that Nature has chosen genetically the best mode then we may propose the idea that the soliton wave becomes a standing one at TSs. By the standing wave, we assume the one whose envelope velocity is equal to the RNAP velocity. As the latter one is negligible we state

$$V_e = 0. \qquad (58)$$

Thus, Eqs. (57) and (58) provide food for thought, and, therefore, must be carefully checked. We, here, perform mathematical analysis of the demodulated standing soliton (DSS) mode. This means that we investigate if there exists a certain value of $ql$ satisfying these equations. To simplify the mathematics, we introduce new parameters $x$, $b$ and $s$ defined through relations

$$K = xk, \qquad b = 1-\eta, \qquad a^2 D = sk \qquad (59)$$

and use $h = 5$, as explained earlier. As the parameter $k$ determines the strong covalent bond we know that both $x$ and $s$ should be much less than one.

It is convenient to introduce the following expressions:

$$\left.\begin{array}{l} f_1 \equiv f_1(ql) = \sin(ql) - 5x\sin(5ql) \\ f_2 \equiv f_2(ql) = \cos(ql) - 25x\cos(5ql) \\ f_3 \equiv f_3(ql) = \sin^2(ql/2) + x\cos^2(5ql/2) \end{array}\right\}. \qquad (60)$$

Hence, the expression

$$V_g = \frac{kl}{m\omega} f_1 \qquad (61)$$

is obvious, while Eqs. (44), (57) and (58) bring about a useful formula

$$V_g = 2P\frac{ql}{l}, \qquad V_g(1-\eta) = \frac{\omega l}{ql}. \qquad (62)$$

A next step is to solve the system (61), (39) and (62). Of course, Eqs. (59) and (60) should be applied. If we eliminate $P$ and $V_g$ we straightforwardly obtain

$$b \equiv b(ql) = \frac{f_1}{qlf_2 - f_1}, \qquad M_o \equiv \frac{m\omega^2}{k} = \frac{ql\, f_1^2}{ql\, f_2 - f_1}, \qquad P = \frac{kl^2 f_1}{2qlm\omega}. \qquad (63)$$

Notice that the expression for $P$ is simpler than the corresponding one in [72], although both are correct. Finally, Eqs. (32), (59) and (63) yield to

$$s \equiv s(ql) = \frac{ql\, f_1^2}{4(ql\, f_2 - f_1)} - f_3. \qquad (64)$$

Also, according to Eqs. (21), (35), (36), (39) and (59), we easily obtain

$$\mu = -\frac{2\alpha s}{s+x}, \qquad \delta = \frac{\alpha s}{M_0 - f_4 - s} \equiv \frac{\alpha s}{\delta_d}, \qquad (65)$$

$$Q = -\frac{2s^2 k^2}{m\omega D}\Phi, \qquad \Phi \equiv \Phi(ql) = \frac{7x - 11s}{s+x} + \frac{9s}{\delta_d}, \qquad (66)$$

where

$$f_4 \equiv f_4(ql) = \sin^2(ql) + x\cos^2(5ql). \tag{67}$$

It was mentioned above that we are looking for possible value(s) of $ql$ that satisfy a couple of requirements. It is convenient to assume the wavelength as an integer of $l$, i.e.,

$$ql = \frac{2\pi}{\lambda}l = \frac{2\pi}{N}. \tag{68}$$

In the examples used for Figs. 6, 10 and 11, this integer was $N=10$. If we assume that $N$ cannot be smaller than six then $ql$ should be less than one. Therefore, the big values for $ql$ are not acceptable. This means that $f_1$, existing in Eq. (60), is positive. In fact, it can be negative but for unacceptable large $x$ only. As $f_1 > 0$ for any small enough $ql$, we conclude that $P > 0$ as well.

There are a couple of requirements that should be satisfied. They are: $b > 0.5$, $s > 0$, $\delta_d \neq 0$ and $\Phi < 0$. The last one is coming from Eqs. (66) and (62), as the parameter $Q$ is positive. Let us take $b$ as an example. Figure 12. shows $b(ql)$ for two values of $x$. We see that there should be

$$\left.\begin{array}{ll} 0.25 < ql < 0.81 & \text{for} \quad x = 1/50 \\ ql < 0.77 & \text{for} \quad x = 1/80 \end{array}\right\}. \tag{69}$$

The part $ql > 0.25$ is coming from the figure $s(ql)$. Also, $\delta_d > 0$ for $ql$ bellow the upper limits indicated in Eq. (69).

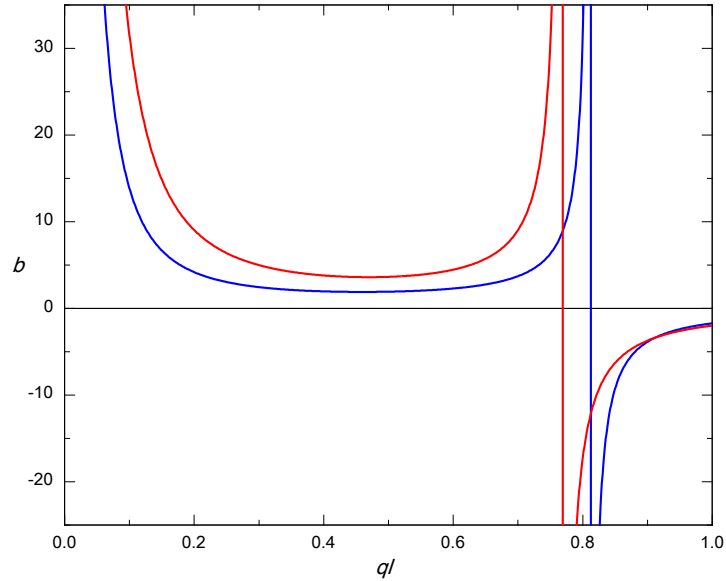

Figure 12: Parameter $b$ as function of $ql$ for: $x = 1/50$ (blue) and $x = 1/80$ (red)

The next step is the function $\Phi(ql)$, existing in Eq. (66). It is shown in Fig. 13. The only conclusion is

$$ql > 0.45 \quad \text{for} \quad x = 1/50. \tag{70}$$

Also, $\delta_d > 0$ in the intervals $ql < 0.81$ ($x = 1/50$) and $ql < 0.77$ ($x = 1/80$), as indicated in Eq. (69). Of course, the final allowed intervals for $ql$ are given by Eqs. (69) and (70).

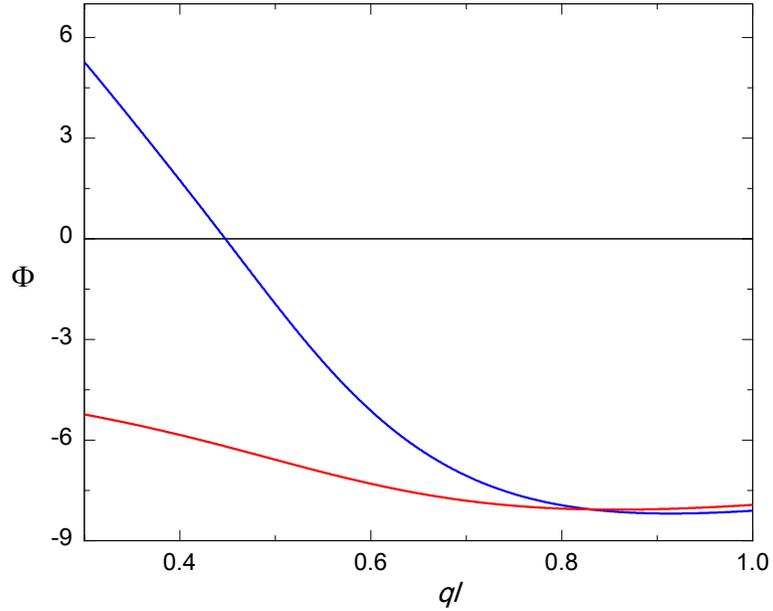

Figure 13: Function $\Phi(ql)$ for: $x = 1/50$ (blue) and $x = 1/80$ (red)

Therefore, there exist the values for $ql$ satisfying the requirements for the DSS mode, i.e., Eqs. (57) and (58). Our final task is to plot the nucleotide pair stretching corresponding to the DSS mode. As an example, we pick $ql = 0.47\text{rad}$ and $x = 1/50$. For $D = 0.07\text{eV}$ and $k = 12\,\text{N/m}$, from Eqs. (43) and (47), we easily calculate $A = 6.1\,\text{Å}$, $\Lambda = 8l$ and $s = 0.03$. The second value means that the wave covers 8 base pairs, which perfectly matches the experimental value for the extent of DNA:RNA hybrid [57]. This soliton is shown in Fig. 14 (blue), together with another example $ql = 0.20\text{rad}$ and $x = 1/80$, which yields to $A = 1.6\,\text{Å}$, $\Lambda = 7.6l$ and $s = 0.05$ (red). The figure obviously shows demodulated solitons. These are nothing but a kind of bell-type solitons. Big amplitudes are in agreement with the local opening of the chain.

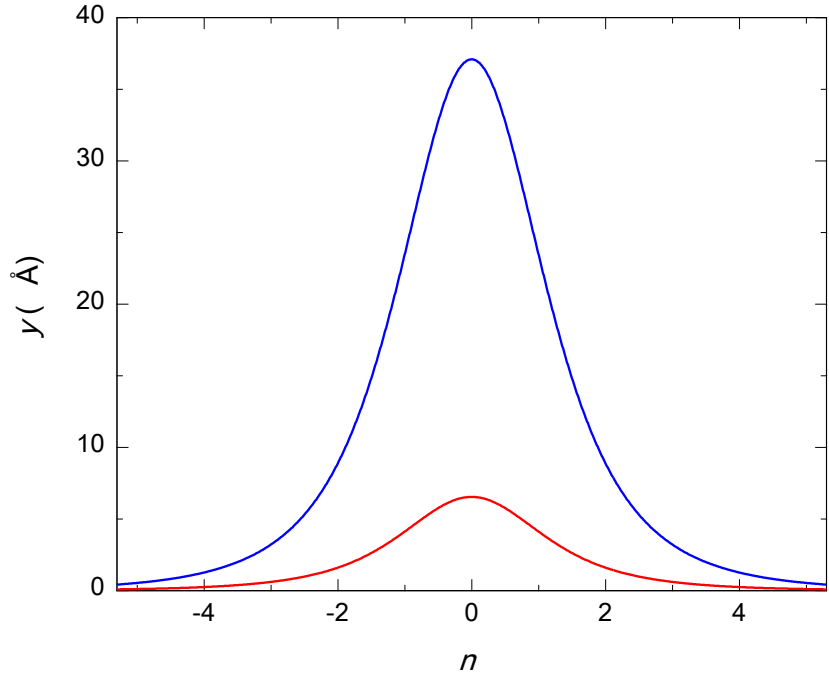

Figure 14: Demodulated solitary waves for: $ql = 0.47\text{rad}$, $x = 1/50$ (blue) and $ql = 0.20\text{rad}$, $x = 1/80$ (red)

Therefore, we demonstrated that $ql$ satisfying our postulates explained above exists. Importantly, for the acceptable values of the relevant parameters, the corresponding soliton width matches the experimental value.

We complete this chapter with a couple of concluding remarks. We have dealt with DNA modeling. Two models, the Y, and HPB, are explained in some more details. They are examples of the angular and radial models, respectively. There have been a variety of attempts to improve these models. One important example is a model representing a combination of the Englander's and PB models, which brings about the kink soliton [76]. Another example is the introduction of asymmetric double Morse potential [77,78], instead of the one used here.

Let us compare the models explained in the first part of the subsection 13.1 and HPB one. A reader might come to the conclusion that the first group of them may yield towards the kink-type solitons, while the HPB is reserved for the breathers. However, this would not be quite correct. Within the HPB model, the semi-discrete approximation has been applied. It has turned out that, according to this mathematical procedure, the breathers move along the chain. However, the continuum approximation can be used as well. Following the same procedure as in Section 11, we obtain the kinks moving along the chain [79]. Besides the kinks, the bell-type solitons were obtained in the case of negligible viscosity. Therefore, the final result does not depend on the studied system only, but on the applied mathematics as well. Of course, a crucial question is which of these solitons really exist in DNA, if any. It was argued that kinks and breathers do not exclude each other and that both solitons play an important role in DNA functioning [79]. Let us keep in mind the DNA–RNA transcription again. From the point of view of a single nucleotide pair, two nucleotides oscillate in a transverse direction around a certain distance. If this distance is big, then the local opening will more likely happen. The kink certainly means a certain step, which can be an increase of the distance around which the nucleotides oscillate. If so, then the kink could be understood as a prerequisite for the breather [79]. Regarding the mathematical methods, it is worth mentioning the method based on Jacobi elliptic functions [80] and fractional Lagrange formalism [81,82]. They yield to the breathers again, i.e., to the results obtained using the semi-discrete approximation and shown above.

It was pointed out that the prerequisite for DNA-RNA transcription is the local opening. However, it is known that DNA must also unwind locally to let one strand serve as a template for the synthesis of a new strand of RNA [5]. The HPB model can take this into consideration. Namely, the local angle of helix winding is defined through the value of the parameter $h$. The partially unwinding chain corresponds to the decrease of $h$. We have been dealing with constant $h$ so far. However, possible generalization, i.e., allowing it to be a variable at TSs, could be a topic of further research.

It might be important to point out that thermodynamics of local DNA opening was studied in [83], relying on the PB model. One of the future research tasks should be an extension of this work. This means that the HPB model should be used instead of its simpler predecessor. The HPB model is doubtlessly better than the PB one and its advantage is especially important when we study the local opening. Namely, the term comprising $K$ is extremely important as it describes helicoidal structure of DNA.

A patient reader has certainly noticed that the big amplitudes are not in agreement with the earlier assumptions of small oscillations. This means that the HPB model only predicts the RM and DSS mode but is not adequate for complete quantitative analysis.

Let us complete this section with a few more words about the HPB model. The RM and DSS mode have been studied neglecting viscosity. Introduction of the dumping effects would be an important advantage but a real challenge as well. Also, these two modes have been studied independently. One of the future tasks should be an attempt to involve both of them into a single theory. It is clear that DNA modeling has accomplished tremendous progress during the past a couple of decades. However, the velocity of increasing the volume of the knowledge cannot match the velocity of appearing new questions. This means that the life of a biophysicist working on this topic is very interesting.